\documentclass[sigconf,nonacm]{acmart}

\setcopyright{none}
\renewcommand\footnotetextcopyrightpermission[1]{}

\usepackage{graphicx}
\usepackage{hyperref}
\usepackage[whole]{bxcjkjatype}
\usepackage{booktabs}
\usepackage{xurl}
\usepackage{amsmath}
\usepackage{appendix}

\usepackage{algpseudocode}

\usepackage{caption}
\usepackage{algorithm}
\usepackage{algpseudocode}

\usepackage{multirow}

\usepackage[most]{tcolorbox}
\tcbuselibrary{breakable,raster,skins}

\usepackage{fancybox}

\usepackage{longtable}
\usepackage{array}

\usepackage{tikz}
\usepackage{eso-pic}

\begin{document}

\AddToShipoutPictureBG*{%
  \AtPageUpperLeft{%
    \begin{tikzpicture}[remember picture, overlay]
      \node[anchor=north, xshift=0.5\paperwidth, yshift=-1cm] {%
        \begin{tcolorbox}[
          colback=white,
          colframe=black,
          arc=0mm,
          boxrule=0.5pt,
          top=2pt,
          bottom=2pt,
          left=4pt,
          right=4pt,
          width=0.9\textwidth
        ]
          This paper has been published in an updated form in IEEE Access as:
\textit{T. Koide, H. Nakano and D. Chiba, "ChatPhishDetector: Detecting Phishing Sites Using Large Language Models," in IEEE Access, vol. 12, pp. 154381-154400, 2024, doi: 10.1109/ACCESS.2024.3483905.}
        \end{tcolorbox}
      };
    \end{tikzpicture}
  }
}

\title{Detecting Phishing Sites Using ChatGPT}

\settopmatter{authorsperrow=4}
\author{Takashi Koide}
\email{takashi.koide@global.ntt}
\orcid{0009-0008-1942-0335}
\affiliation{%
  \institution{NTT Security Japan (KK)}
  \city{Tokyo}
  \country{Japan}
}
\author{Naoki Fukushi}
\email{naoki.fukushi@global.ntt}
\affiliation{%
  \institution{NTT Security Japan (KK)}
  \city{Tokyo}
  \country{Japan}
}
\author{Hiroki Nakano}
\email{hi.nakano.sec@gmail.com}
\affiliation{%
  \institution{NTT Security Japan (KK)}
  \city{Tokyo}
  \country{Japan}
}
\author{Daiki Chiba}
\email{daiki.chiba@ieee.org}
\orcid{0000-0002-7532-6633}
\affiliation{%
  \institution{NTT Security Japan (KK)}
  \city{Tokyo}
  \country{Japan}
}
\author{}

\renewcommand{\shortauthors}{Koide et al.}

\begin{abstract}
The emergence of Large Language Models (LLMs), including ChatGPT, is having a significant impact on a wide range of fields.
While LLMs have been extensively researched for tasks such as code generation and text synthesis, their application in detecting malicious web content, particularly phishing sites, has been largely unexplored. 
To combat the rising tide of cyber attacks due to the misuse of LLMs, it is important to automate detection by leveraging the advanced capabilities of LLMs.

In this paper, we propose a novel system called \textsc{ChatPhishDetector} that utilizes LLMs to detect phishing sites. Our system involves leveraging a web crawler to gather information from websites, generating prompts for LLMs based on the crawled data, and then retrieving the detection results from the responses generated by the LLMs.
The system enables us to detect multilingual phishing sites with high accuracy by identifying impersonated brands and social engineering techniques in the context of the entire website, without the need to train machine learning models.
To evaluate the performance of our system, we conducted experiments on our own dataset and compared it with baseline systems and several LLMs.
The experimental results using GPT-4V demonstrated outstanding performance, with a precision of 98.7\% and a recall of 99.6\%, outperforming the detection results of other LLMs and existing systems. 
These findings highlight the potential of LLMs for protecting users from online fraudulent activities and have important implications for enhancing cybersecurity measures.

\end{abstract}



\maketitle

\section{Introduction}

Large Language Models (LLMs)~\cite{gpt,gpt-3,instructgpt,gpt-4}, such as ChatGPT~\cite{chatgpt}, are revolutionizing various domains, including natural language understanding, generation, and interaction. 
While previous research has focused on exploring the capabilities of LLMs for tasks such as code generation and text synthesis, the potential of LLMs in analyzing and detecting malicious web content, particularly phishing sites, remains unexplored.
Phishing sites pose a severe threat to Internet users. They employ social engineering (SE) techniques~\cite{YangAL23,NelmsPAA16,VadrevuP19} and masquerade as legitimate platforms, tricking users into revealing sensitive information or causing financial harm.
To address the growing threat of automated cyber attacks, including those involving LLMs~\cite{lin2024malla,GuptaAAPP23,GadyatskayaSpringer23,NaitoIEEE23}, it is important to automate the detection of malicious web content by leveraging the high versatility of LLMs.

In this study, we propose a novel system called \textsc{ChatPhishDetector} to detect phishing sites using LLMs. 
Our system employs a web crawler to collect website information and generate prompts~\cite{Wei0SBIXCLZ22,KojimaGRMI22} for LLM analysis. This approach allows for the identification of SE techniques and brand impersonation without requiring extensive training data for machine learning models.
To the best of our knowledge, this is the first study to analyze the ability of LLMs to identify phishing sites.
To evaluate the performance of our system, we conducted a comparison experiment between several LLMs and the existing systems for phishing site detection.
The experimental results using GPT-4 with vision (GPT-4V)~\cite{gpt4v} showed outstanding performance, with a precision of 98.7\% and a recall of 99.6\%.
We found that GPT-4/GPT-4V excelled in its ability to assess the suspiciousness of domain names, identify SE techniques from the website content, and provide comprehensive phishing detection results by considering multiple factors.
The results of this study emphasize the potential of LLMs in efficiently detecting phishing sites, particularly in uncovering SE techniques aimed at psychologically manipulating users.
These findings have significant implications for enhancing automated cybersecurity measures and mitigating the risks of online fraudulent activities faced by users.

\begin{figure*}[!t]  
  \centering  
  \includegraphics[width=0.8\linewidth]{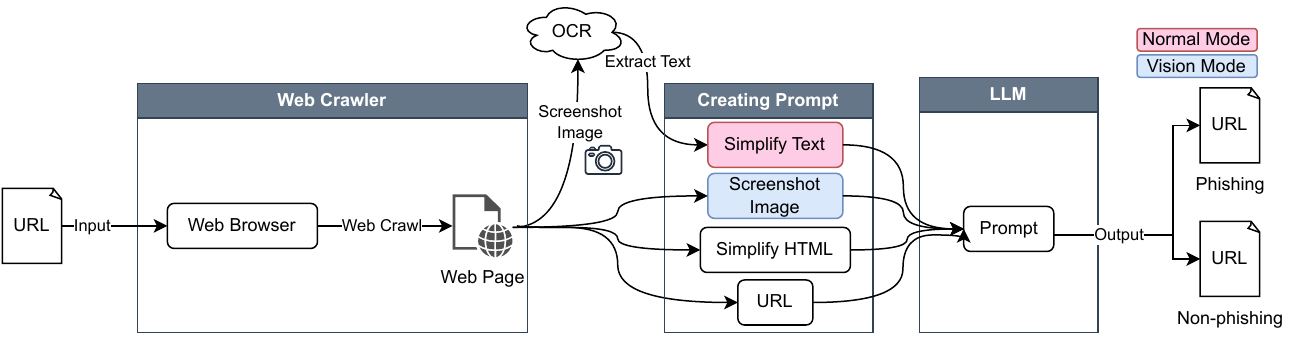}  
  \caption{Overview of \textsc{ChatPhishDetector}.}  
  \label{fig:overview}  
\end{figure*}

In summary, we make the following contributions:
\begin{itemize}
\item We propose a novel system called \textsc{ChatPhishDetector} for detecting phishing sites using LLMs. Our system automatically classifies whether a website is phishing or not by generating prompts based on web crawling results.
\item We conduct an experimental evaluation of our system, comparing it with several LLMs and existing systems, and show that GPT-4V exhibited the highest precision at 98.7\% and recall at 99.6\% in identifying phishing sites.
\item We present a detailed analysis of responses generated by LLMs and explore why LLMs exhibit advanced capabilities in analyzing phishing sites. Our findings revealed that GPT-4/GPT-4V excel in identifying suspicious domain names associated with brand impersonation, finding SE techniques employed in web content, and prioritizing multiple factors for comprehensive phishing detection.
\end{itemize}

\section{Background}
Phishing sites are fraudulent websites that aim to steal personal information or money, or to cause malware infections by psychologically luring users. Attackers use email, short message service (SMS), and web advertisements to attract users and redirect them to phishing sites by having them click on malicious links~\cite{TangML0022,SzurdiLKNC21}. There are two key components of phishing sites: (1) imitation of legitimate services using official logos and branding, and (2) use of social engineering techniques to manipulate user actions. The phishing sites targeted in this study have one or both of those components.
By imitating legitimate services, phishing sites deceive users and gain their trust~\cite{LinLDNCLSZD21,Liu0YNDD22}. These websites often impersonate legitimate platforms such as online banking~\cite{BijmansBSNW21}, e-commerce sites~\cite{BitaabCO23}, and social media~\cite{ThomasLZBRIMCEM17}. In addition, they can create a sense of urgency or fear by displaying fake malware infection warnings or account problems, or generate interest by displaying fake rewards~\cite{KoideDIMVA2020}. Through these SE techniques, users are misled into providing sensitive information such as login credentials and credit card numbers. Furthermore, various user actions may be induced, such as sending cryptocurrency, calling fake technical support centers~\cite{MiramirkhaniSN17}, or downloading apps and executable files~\cite{NelmsPAA16}.

Previous studies have been conducted to understand the characteristics, techniques, and fundamental mechanisms used by attackers in phishing sites. Researchers analyzed the design, structure, and content of these sites, identified common patterns, and developed methods for detecting phishing sites. These studies can contribute to improving security practices, educating users about potential threats, and devising effective strategies to mitigate online fraud and risks. For example, some studies exist on identifying websites that abuse legitimate branding based on their appearance~\cite{AbdelnabiKF20,LinLDNCLSZD21, Liu0YNDD22}, as well as on discovering brand information in domain names and URLs~\cite{AgtenJPN15,NikiforakisAMDPJ13,KintisMLCGPNA17}. Other studies have also been conducted on identifying phishing sites based on information contained in certificates and domain names~\cite{DrichelDBM21,KimCKDSAD21}, and on capturing the context of social engineering in technical support scams, fake infection warnings, and fake rewards~\cite{KharrazRK18,Koide0A20,KoideDIMVA2020}.

Although various methods have been employed to detect phishing sites, there are two main problems. One is the need for learning targeted brands and modify detection logic accordingly for phishing sites. 
For example, it is necessary to collect in advance the logo images that phishing sites abuse, or to create rules to detect them according to domain squatting techniques.
The other is the inability to analyze in detail the context of the psychological manipulation induced by SE techniques. While some studies have been conducted on keyword matching and deep learning-based analysis~\cite{KharrazRK18,Yang2019PhishingWD}, no attempt has been made to automatically analyze and understand the various contexts of psychological manipulation by examining the entire content of a website.

\section{\textsc{ChatPhishDetector}}
We propose \textsc{ChatPhishDetector} a system that uses LLMs to analyze the content of websites and URLs to detect phishing sites. Our system takes advantage of LLMs' highly accurate contextual understanding to precisely identify textual and visual representations involving SE techniques, as well as inconsistencies between brands deceived by the websites and domain names.
By employing LLMs, which is pre-trained on extensive data, our system can detect various phishing sites in multiple languages without learning from collected phishing sites. In this manner, the system addresses two problems present in previous studies. An overview of the system is illustrated in Figure~\ref{fig:overview}.
Our system uses a web crawler to access the input URL and obtain information from the visited website, such as a screenshot image, HTML, and the URL. This information is used to create a prompt for input into LLMs, which then determines whether the website is a phishing site or a non-phishing site.
Our system has two modes: \textit{Normal mode}, which operates with LLMs that accept text input only, and \textit{Vision mode}, which operates with multimodal LLMs that accept both text and images as input.

\subsection{Web Crawling}
\label{sec:webcrawling}

We implement a web crawler that automates Google Chrome using Chrome DevTools Protocol~\cite{cdp} to visit websites and collect information.
Given an input URL, the web crawler retrieves the URL of the reached web page (the final destination after any redirects), the HTML after JavaScript execution, and captures a screenshot image.
The reason for obtaining the HTML after JavaScript execution (Browser-rendered HTML) is that some phishing sites employ obfuscated JavaScript to generate DOM elements as a means to evade analysis. As a result, it becomes challenging to analyze any traces of phishing solely based on the HTML before JavaScript execution.
The web crawler is configured to emulate two different environments: Google Chrome on Windows and Safari on iPhone. These configurations included specifying the \textit{UserAgent} and browser size for each environment.

\subsection{Prompt Engineering}

\definecolor{highlightcolor}{RGB}{255,0,0}

\newcommand{\hnormal}[1]{%
  \tikz[baseline=(X.base)] \node[rectangle, fill=magenta!30, inner sep=2pt] (X) {#1};%
}
\newcommand{\hvision}[1]{%
  \tikz[baseline=(X.base)] \node[rectangle, fill=blue!30, inner sep=2pt] (X) {#1};%
}
\newcommand{\hinput}[1]{%
  \tikz[baseline=(X.base)] \node[rectangle, fill=gray!50, inner sep=2pt] (X) {#1};%
}

\newtcbtheorem[auto counter]{prompt}{Prompt Template}{}{prom}
\begin{figure}[!t]
\begin{prompt}[fontupper=\small\ttfamily]{}{}
\textcolor{gray}{\textnormal{Note: \hnormal{Normal mode} is indicated with a red marker, and \hvision{Vision mode} is indicated with a blue marker.}}\\
You are a web programmer and security expert tasked with examining a web page to determine if it is a phishing site or a legitimate site. To complete this task, follow these sub-tasks:\\\vspace{-7pt}
\\
1. Analyze the HTML, URL, and \hnormal{OCR-extracted text} \hvision{screenshot image} for any SE techniques often used in phishing attacks. Point out any suspicious elements found in the HTML, URL, or text.\\
2. Identify the brand name. If the HTML appears to resemble a legitimate web page, verify if the URL matches the legitimate domain name associated with the brand, if known.\\
3. State your conclusion on whether the site is a phishing site or a legitimate one, and explain your reasoning. If there is insufficient evidence to make a determination, answer "unknown".\\
4. Submit your findings as JSON-formatted output with the following keys:\\
- phishing\_score: int (indicates phishing risk on a scale of 0 to 10)\\
- brands: str (identified brand name or None if not applicable)\\
- phishing: boolean (whether the site is a phishing site or a legitimate site)\\
- suspicious\_domain: boolean (whether the domain name is suspected to be not legitimate)\\\vspace{-7pt}
\\
Limitations:\\
- The HTML may be shortened and simplified.\\
\hnormal{- The OCR-extracted text may not always be accurate.}\\\vspace{-7pt}
\\
Examples of social engineering techniques:\\
- Alerting the user to a problem with their account\\
- Offering unexpected rewards\\
- Informing the user of a missing package or additional payment required\\
- Displaying fake security warnings\\\vspace{-7pt}
\\
URL:\\
\hinput{\{URL\}}\\\vspace{-7pt}
\\
HTML:\\
\verb|```|
\hinput{\{Browser-rendered HTML\}}
\verb|```|\\\vspace{-7pt}
\\
\hnormal{Text extracted using OCR:}\\
\hnormal{\texttt{\`{}\`{}\`{}}
\hinput{\{OCR-extracted text\}}
\texttt{\`{}\`{}\`{}}}
\end{prompt}
\end{figure}

We describe the process of generating a prompt for detecting phishing sites.
The template for the prompt is shown in Prompt Template 1.
The purpose of this prompt is to provide website information to LLMs and control the inference process to determine whether it is a phishing site or not.
We create the template based on the Chain of Thought (CoT) prompting technique~\cite{wei2023chainofthought,KojimaGRMI22}.
The CoT is a prompting technique that encourages LLMs to explain their reasoning. 
The CoT has been shown to be effective in improving performance on a variety of reasoning tasks, such as arithmetic and symbolic reasoning.
The task of phishing site detection is divided into four subtasks, which facilitate the execution of specific reasoning processes. These subtasks are as follows:
\begin{enumerate}
\item Analyze whether the website contains SE techniques that deceive or attract users. The prompt provides typical SE techniques commonly used by phishing sites, such as cash prizes, fake malware infection warnings, account problems, and postal parcel issues. We instruct LLMs that the presence of these factors without context in a non-login state indicates a high probability of a phishing site.
\item Extract the brand name of the website. Phishing sites may be created by copying resources such as HTML and images from legitimate sites, making it difficult to determine their authenticity based on HTML alone. Therefore, LLMs are prompted to confirm whether the URL corresponds to the legitimate one.
\item Determine whether the website is a phishing site or not, and explicitly state the rationale for the decision. We intend that providing a detailed explanation will improve the accuracy of the response and facilitate human analysis.
\item Generate output in JSON format. If the website uses SE techniques, LLMs are expected to return \textit{phishing} as true. If the LLM identifies the brand name and confirms that its domain name is different from the legitimate one, \textit{suspicious\_domain} is expected to be true. In addition, a \textit{phishing\_score} is generated, which ranges from 0 to 10.
\end{enumerate}
The system creates a prompt by substituting (\romannumeral 1) URL, (\romannumeral 2) HTML, and (\romannumeral 3) text extracted from a screenshot image using optical character recognition (OCR) into this prompt template.
Some phishing sites evade HTML-based phishing site analysis by not including specific brand names or text in the HTML. For example, they display brand information (logos or headings) in image files or canvas elements.
To incorporate such information into the prompt, we employ text extraction from screenshot images using OCR.
In Vision mode, the system sends screenshot images directly to LLMs instead of OCR-extracted text.
To extract detection results from the responses, we define a response as phishing if either the value of the \textit{phishing} key or the \textit{suspicious\_domain} key is true. Conversely, a response is labeled as non-phishing if both keys are false.

Depending on the length of HTML and OCR-extracted text, some websites may exceed LLMs' token (characters or words used for processing by LLMs) limit.
For example, GPT-3.5-turbo and Llama 2 have a 4k token limit. In fact, the median token count for HTML in our dataset (see Section~\ref{sec:dataset}) is 30,398 (with a median of 74,937 for non-phishing pages), indicating that a large number of web pages exceed this token limit. To compare the performance of different LLMs under the same conditions, we simplify the HTML and OCR-extracted text to fit within 4k tokens in this study. %
Since this prompt template consumes 362 tokens, we keep 300 tokens for the response and set the maximum tokens for HTML, OCR-extracted text, and URL to 2,500, 500, and 300, respectively.
In the following sections, we describe the specific processes to simplify the HTML and OCR-extracted text.

\subsection{Simplifying HTML}
\label{sec:simpifyingHTML}

The system simplifies the HTML while preserving essential information for phishing detection and brand identification. Elements located at the top of the HTML, such as the \texttt{title} and the \texttt{meta description} element, primarily contain cues for determining the services provided by the website. Additionally, the \texttt{form} element used for inputting login information and text displayed on the page are crucial for analyzing the role of the website. To retain the amount of such information while minimizing the overall HTML length, we follow the steps outlined in Algorithm~\ref{alg:html-processing}.
This process removes HTML elements, such as style, script, and comment tags, that have low relevance and a high number of tokens. Subsequently, the system unwraps HTML elements other than the important tags listed as \texttt{head, title, meta, body, h1, h2, h3, h4, h5, h6, p, strong, a, img, hr, table, tbody, tr, th, td, ol, ul, li, ruby, and label}. Unimportant elements are eliminated, and their child elements are incorporated into their parent elements.
Then, our system removes HTML elements that do not contain text enclosed within tags. We also shorten the \texttt{src} attribute \texttt{img} elements encoded in base64 and the \texttt{href} attribute of \texttt{a} elements containing lengthy URLs. The system repeatedly removes an HTML element located in the middle of the HTML until the number of tokens falls below the maximum number of tokens.

\begin{figure}[!t]
\centering
\begin{algorithm}[H]
\begin{algorithmic}[1]
\caption{Simplifying HTML}
\label{alg:html-processing}
\Require{$\text{inputHTML}$: Input HTML}
\Ensure{$\text{processedHTML}$: Processed HTML}
\Function{SimplifyHTML}{$\text{inputHTML}$}
\State Remove style, script, and comment elements from $\text{inputHTML}$ \label{step:remove-elements}
\State $\text{processedHTML} \gets \text{Result after removal}$

\If{$\text{lengthToken}(\text{processedHTML}) < 3000$} \label{step:length-check-1}
\State \Return $\text{processedHTML}$
\EndIf

\State Unwrap elements except for important tags \label{step:unwrap-elements}
\State Remove elements without text content \label{step:remove-empty-elements}
\State Shorten \texttt{href} in \texttt{a} tags and \texttt{src} in \texttt{img} tags \label{step:shorten-attributes}
\State $\text{processedHTML} \gets \text{Result after removal}$

\If{$\text{lengthToken}(\text{processedHTML}) < 3000$} \label{step:length-check-2}
\State \Return $\text{processedHTML}$
\EndIf

\While{$\text{lengthToken}(\text{processedHTML}) > 3000$} \label{step:length-check-loop}
\State Remove an HTML element from the midpoint of the processedHTML \label{step:reduce-token-count}
\State $\text{processedHTML} \gets \text{Result after removal}$
\EndWhile

\State \Return $\text{processedHTML}$
\EndFunction
\end{algorithmic}
\end{algorithm}
\end{figure}

\subsection{Simplifying OCR-extracted Text}
\label{sec:simpifyingOCR}

Our system also simplifies OCR-extracted text in normal mode.
Specifically, it examines the font size of the identified text by the OCR and removes sentences starting from the smallest font size.
This process is repeated until the number of tokens in the text falls below the maximum number of tokens.

\section{Dataset}
\label{sec:dataset}

We describe how we created the dataset used in our experiment.
The dataset includes a total of 1,000 phishing sites and an equal number of non-phishing sites.
To gather these websites, we conducted web crawling starting from the collected seed URLs.

\subsection{Phishing Sites}
\label{sec:dataset_phishing}

To collect phishing sites, we used OpenPhish~\cite{openphish} and PhishTank~\cite{phishtank}, which are phishing intelligence sources, and CrowdCanary~\cite{nakano2023canary}, a method for extracting phishing-related posts from Twitter using machine learning models. CrowdCanary uses keyword-based filters such as ``phishing'' and ``scam'' and extracts URLs from posts associated with phishing encountered by victims or observed by security experts. From March to April 2023, we collected URLs from these three sources and used them as seed URLs to crawl.
This enabled us to obtain a wide range of phishing sites that are not limited to incoming channels such as email, SMS, and web advertisements. 
Subsequently, we accessed the seed URLs using the web crawler described in Section~\ref{sec:webcrawling}. 
To ensure the integrity of our dataset, we excluded websites with incomplete rendering and image loading. 
We then selected phishing sites that contained one of the following two elements as candidates for the dataset.
\begin{itemize}
\item Websites that display logos or brand names associated with well-known services or that visually resemble legitimate sites.
\item Websites employing SE techniques, such as presenting false information (e.g., fake rewards, virus infection alerts, account issues), to generate a sense of urgency or interest.
\end{itemize}

\begin{figure}[!t]  
  \centering  
  \includegraphics[width=0.9\linewidth]{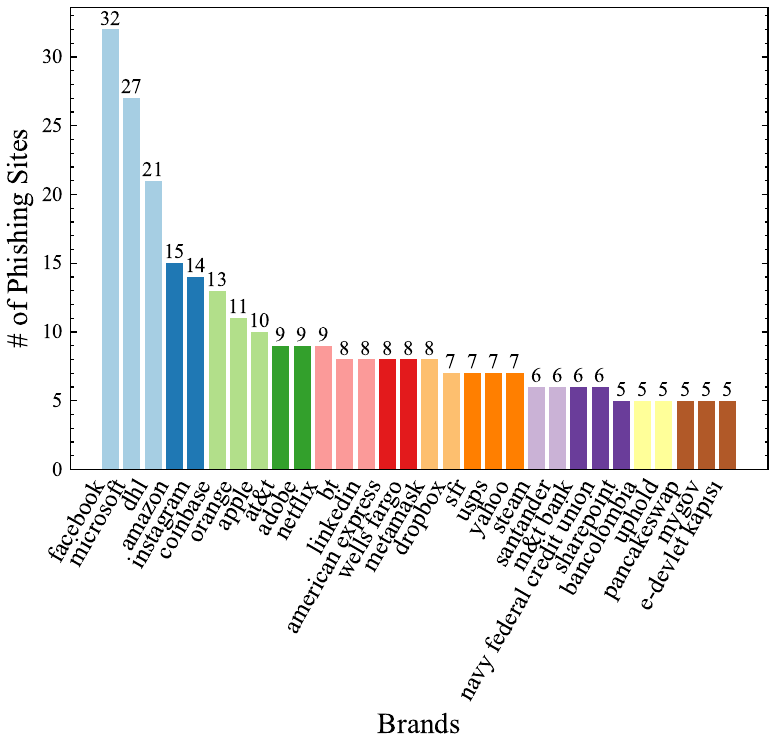}  
  \caption{Top 30 Brands Targeted by Phishing Sites.}  
  \label{fig:brands}  
\end{figure}  

Where similar phishing sites were found among the extracted phishing sites, we excluded all but one. Criteria for determining similarity include fully qualified domain name (FQDN), page title, website appearance (calculated in perceptual hash), and text content. However, if the websites had the same appearance but different languages, they were retained.

\begin{figure}[!t]  
  \centering  
  \includegraphics[width=0.85\linewidth]{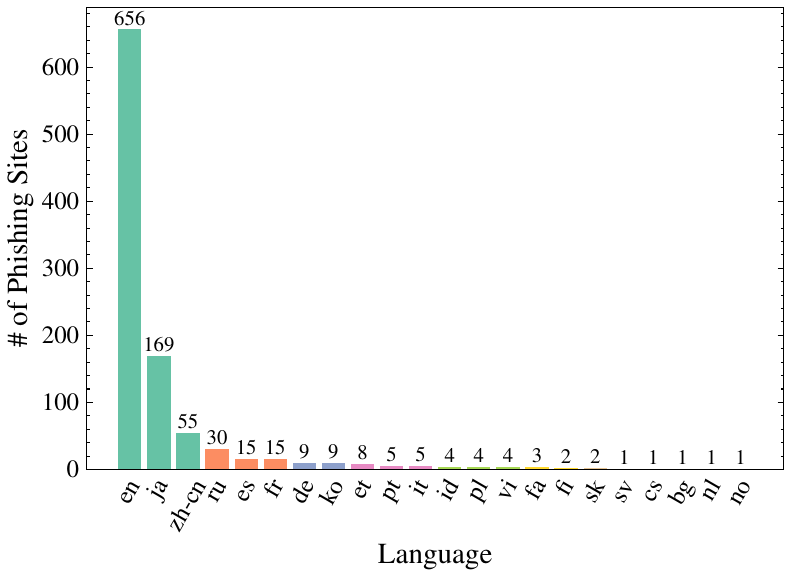}  
  \caption{Language Distribution of Phishing Sites.}  
  \label{fig:language_phish}  
\end{figure}

As a result of the above analysis, we obtained a dataset of 1,000 phishing sites with unique 1,000 FQDNs. These phishing sites targeted a total of 147 legitimate service brands (see Appendix~\ref{appendix_brands}). We identified 32 phishing sites that did not contain any brand information.
Figure~\ref{fig:brands} shows the top 30 brands used by phishing sites in the dataset.
We extracted text from the screenshot images using OCR for the normal mode. In this paper, we used Azure Cognitive Service~\cite{azurecognitive} to streamline the OCR process for multilingual websites. We identified the languages of the OCR-extracted text using langdetect~\cite{langdetect}, a language detection library, revealing that the 1,000 phishing sites were distributed across 22 different languages. Figure~\ref{fig:language_phish} illustrates the number of phishing sites in each language.

Phishing sites vary in design; some are exact clones of legitimate sites, while others have original content or incorporate elements such as layouts and logos from legitimate sites. These diverse phishing sites are difficult to detect using existing systems that rely primarily on image similarity to their legitimate sites. We identified 172 phishing sites (17.2\%) in our dataset that did not closely resemble their legitimate sites.
Examples of such phishing sites include those that mimic popular platforms such as Amazon or Netflix and employ gift card scams. There are also phishing sites that impersonate Facebook logos or disguise Windows Defender scan results, displaying fake account suspensions or virus infection warnings.

\begin{figure}[!t]  
  \centering  
  \includegraphics[width=0.85\linewidth]{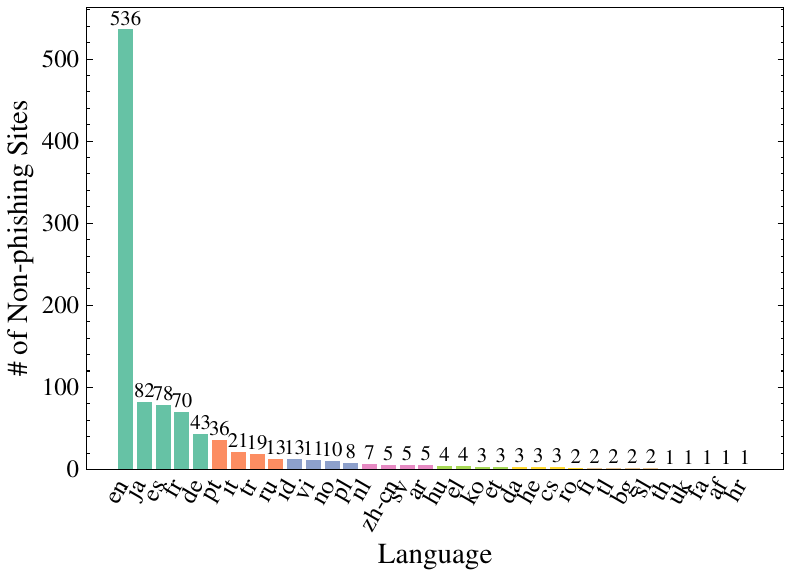}  
  \caption{Language Distribution of Non-phishing Sites.}  
  \label{fig:language_ben}  
\end{figure}

\subsection{Non-phishing Sites}
We collected an equal number of websites as phishing sites for non-phishing sites.
The seed URLs consisted of legitimate websites for the 153 brands targeted by phishing sites, as well as the top 2k domain names from the Tranco list~\cite{PochatGTKJ19tranco}.
From the legitimate sites of the 153 brands, we extracted the URLs of their homepages and, if available, the login pages (a total of 236 URLs with 196 FQDNs).
After crawling the Tranco top 2k, we successfully accessed 1,661 URLs.
The Tranco top sites include certain categories such as adult content, illegal downloads/streaming, and gambling. Websites within these categories may not always be benign, as they potentially lead to phishing sites through malicious advertising. Consequently, we have excluded such websites from our analysis.
The breakdown of excluded websites is as follows: 28 porn, 3 gambling, and 9 illegal sites.
The remaining 764 websites (out of 1,000 non-phishing sites) were randomly selected from the crawled Tranco top sites.
The non-phishing sites are distributed across 34 languages.
Figure~\ref{fig:language_ben} presents the distribution of non-phishing sites across different languages.

\subsection{Simplifying HTML and OCR-Extracted Text}

We simplified the collected HTML and OCR-extracted text of phishing sites and non-phishing sites using the process explained in Section~\ref{sec:simpifyingHTML} and~\ref{sec:simpifyingOCR}.
The median token count in the HTML of non-phishing sites was 74,937 before simplification, while it was 8,275 for phishing sites. 
In general, non-phishing sites have a higher token count. 
One reason for this difference is that non-phishing sites often implement a variety of sophisticated features through complex JavaScript code or use large platforms, while phishing sites have only the minimum functionality necessary to deceive users or steal sensitive information.
Of the 2,000 sites in the data set, 980 exceeded the maximum token limit of 32K tokens for the ChatGPT model used in this study.
The median number of tokens in the OCR-extracted text from non-phishing websites was 296, while for phishing websites, it was 122.
Similar to the HTML comparison, non-phishing sites tend to have a higher number of tokens.

\section{Evaluation}
To validate the detection accuracy of \textsc{ChatPhishDetector}, we performed experiments with different LLMs, different modes, and compared with existing systems.

\begin{table*}[]
\setlength{\tabcolsep}{0.9pt} %
\centering
\caption{Comparison of Performance Metrics. URL, HTML, and Image Indicate the Types of Information Accepted as Input, While Phishing and Non-phishing Represent Classification Targets.}
\label{tab:performance_metrics}
{\renewcommand\arraystretch{0.75}
\begin{tabular}{l|ll|rrrr|ccc|cc}
\toprule
        \textbf{System} & \textbf{Mode} & \textbf{Model} & \textbf{Precision} & \textbf{Recall} & \textbf{Accuracy} & \textbf{F-measure} & \textbf{URL} & \textbf{HTML} & \textbf{Image} & \textbf{Phishing} & \textbf{Non-phishing}  \\
        \midrule
        \textsc{ChatPhishDetector} & Vision & GPT-4V & \textbf{98.7}\% & \textbf{99.6}\% & \textbf{99.2}\% & \textbf{99.2}\% & \checkmark & \checkmark & \checkmark & \checkmark & \checkmark  \\ 
         &  & Gemini Pro Vision & 78.9\% & 99.1\% & 89.1\% & 87.9\% & \checkmark & \checkmark & \checkmark & \checkmark & \checkmark  \\ 
         & Normal & GPT-4 & 98.3\% & 98.4\% & 98.4\% & 98.4\% & \checkmark & \checkmark & \checkmark & \checkmark & \checkmark  \\ 
         &  & GPT-3.5 & 98.3\% & 86.7\% & 92.6\% & 92.1\% & \checkmark & \checkmark & \checkmark & \checkmark & \checkmark  \\ 
         &  & Llama-2-70B & 78.4\% & 66.4\% & 74.1\% & 71.9\% & \checkmark & \checkmark & \checkmark & \checkmark & \checkmark  \\ 
         &  & Gemini Pro & 90.5\% & 95.6\% & 93.2\% & 93.0\% & \checkmark & \checkmark & \checkmark & \checkmark & \checkmark  \\ 
         & Simple & GPT-4 & 98.4\% & 75.5\% & 87.2\% & 85.5\% & \checkmark &  &  & \checkmark & \checkmark  \\ 
         &  & GPT-3.5 & 98.6\% & 77.5\% & 88.2\% & 86.8\% & \checkmark &  &  & \checkmark & \checkmark \\ 
        dnstwist~\cite{dnstwist} & - & - & 31.3\% & - & - & - & \checkmark &  &  & \checkmark &  \\ 
        Phishpedia~\cite{LinLDNCLSZD21} & - & - & 26.0\% & - & - & - &  &  & \checkmark & \checkmark &  \\ 
\bottomrule
\end{tabular}
}
\end{table*}

\subsection{Experimental Setup}
\label{sec:experimental_setup}
In this section, we provide an explanation of LLMs and existing phishing detection systems used in our evaluation experiments.

\subsubsection{Models}
\noindent\textbf{ChatGPT} is the most well-known LLM developed by OpenAI. We utilize the Azure OpenAI API~\cite{azureopenai} for accessing \texttt{gpt-4-0314} (referred to as GPT-4) and \texttt{gpt-3.5-turbo-0301} (referred to as GPT-3.5) in normal mode, and \texttt{gpt-4-vision-preview} (referred to as GPT-4V) in vision mode.

\noindent\textbf{Gemini Pro} is an LLM developed by Google DeepMind~\cite{gemini}. We accessed it via the Google Cloud API, using Gemini Pro in normal mode and Gemini Pro Vision in vision mode.

\noindent\textbf{Llama 2} is an open-source LLM, and we chose to use its most high-performance model (\texttt{Llama-2-70b-chat-hf}). 
We ran the Llama 2 model on Ubuntu Server 20.04 LTS and Nvidia A100 with 4-bit quantization.
We queried Llama 2 with the same prompts as the other models in normal mode.

\noindent\textbf{Simple mode} uses a simplified prompt with only the URL assigned, to compare with the well-designed prompts in our system. The simplified prompt is demonstrated in Prompt Template 2 in Appendix~\ref{appendix_template}. We conducted experiments using both GPT-4 and GPT-3.5 for the simple Mode.

\subsubsection{Baseline Systems}

We chose dnstwist~\cite{dnstwist} and Phishpedia~\cite{LinLDNCLSZD21} as baseline systems because they have been widely used in previous studies~\cite{Liu0ZLD23,Liu0YNDD22,KimCKDSAD21,TianJ0Y018,KoideFN023} for phishing detection.
Since these systems determine legitimate sites by matching against prepared URL allow lists, we compared only the ability of these systems to predict phishing sites (true positives and false negatives).

\noindent\textbf{dnstwist~\cite{dnstwist}} is an open-source tool designed to generate and validate domain names similar to legitimate domain names that may be abused for phishing sites, such as typo-squatting and brand impersonation. We generated such domain name strings using domain names from non-phishing sites in our dataset as seeds. We compared the strings, excluding the TLDs, to the domain names of the phishing sites in our dataset. If there was a partial match, we considered the phishing detection successful.

\noindent\textbf{Phishpedia~\cite{LinLDNCLSZD21}} is an open-source tool to identify phishing sites from screenshot images by training on brand logo images. To train the model, we added additional brand logo images from the non-phishing sites in our dataset to the publicly available logo images in Phishpedia's code. We used Phishpedia to analyze screenshot images of phishing sites in our dataset. If a brand logo image was detected, we considered the phishing detection successful.

\subsection{Summary of Result}
\label{sec:summary_of_result}

 \newcolumntype{C}[1]{>{\centering\arraybackslash}p{#1}}
 \newcolumntype{L}[1]{>{\raggedright\arraybackslash}p{#1}}
 \newcolumntype{R}[1]{>{\raggedleft\arraybackslash}p{#1}}

\noindent\textbf{Performance Comparison}
Table~\ref{tab:performance_metrics} provides a comparative analysis of the performance metrics for the \textsc{ChatPhishDetector} across various models (GPT-4V/4/3.5, Llama 2, and Gemini Pro/Pro Vision), modes (vision, normal, and simple), and baseline systems (dnstwist and PhishPedia). 
Precision ($ = \frac{TP}{TP + FP}$), Recall ($ = \frac{TP}{TP + FN}$), Accuracy ($ = \frac{TP + TN}{TP + TN + FP + FN}$), and F-measure ($ = 2 \times \frac{Precision \times Recall}{Precision + Recall}$) were used as performance metrics. 
Detailed detection results are presented in Table~\ref{tab:chatphishdetector_result} in Appendix~\ref{sec:appendix_detection_results}.
GPT-4V (vision mode) demonstrated remarkable performance in classifying both phishing and non-phishing sites, achieving a precision of 98.7\% and a recall of 99.6\%, with only 4 false negatives and 13 false positives. GPT-4 (normal mode) also outperformed the other models with 98.3\% accuracy and 98.4\% recall.
The classification performance of GPT-4 and GPT-3.5 for non-phishing sites was comparable, with low false positives, and both had 98.3\% precision.
However, for phishing site classification, GPT-4 significantly improved with a recall of 98.4\% compared to GPT-3.5's 86.7\%. 
Interestingly, the vision mode did not always lead to higher accuracy than the normal mode as in GPT-4V and GPT-4.
Gemini Pro Vision had a lower accuracy of 89.1\% compared to Gemini Pro, which achieved 93.2\% accuracy.
Despite Gemini Pro Vision having fewer false positives (7), its substantial number of false negatives (211) indicated a bias towards positive responses.
One possible reason for this performance gap is that Gemini Pro Vision did not excel at recognizing non-English text from images.
Furthermore, the ability to integrate visual and textual data plays a critical role in accurately identifying phishing attempts.
Llama 2, with an accuracy of 74.1\%, was less effective compared to other models in the normal mode. This may be attributed to insufficient training data and model performance, particularly considering Llama 2's primary training on English data and its limited capability in interpreting non-English text and identifying non-English brands.
A comparison between normal and simple modes revealed a significant improvement in accuracy for both GPT-4 and GPT-3.5. 
This improvement is attributed to the normal mode's use of well-considered custom prompts to analyze a combination of URLs, HTML, and screenshot images. It enhances the ability to distinguish phishing sites compared to the simple mode, which relies solely on URLs as input.
In Section~\ref{sec:dataset_phishing}, we discussed the phishing sites in the dataset that do not closely resemble legitimate sites. When analyzing the true positives in this data, GPT-4/GPT-4V successfully identified all 172 phishing sites (100\%), while GPT-3.5 detected 171 phishing sites (99.4\%). Our system demonstrated high accuracy in detecting these phishing sites by analyzing the context of brand impersonation and social engineering techniques, rather than relying solely on comparisons with legitimate sites using screenshot images.

\noindent\textbf{Baseline Systems}
Since dnstwist and Phishpedia focus exclusively on the prediction of phishing sites, we only calculated the precision for these systems. The results showed a precision of 31.3\% for dnstwist and 26.0\% for Phishpedia. For example, dnstwist successfully detected phishing sites using the domain squatting technique such as \texttt{microsft} (a transformation of \texttt{microsoft}) in \url{exceklcmicrosftprotection.pages[.]dev}. Phishpedia is useful for detecting phishing sites with logo images, even if the page design differs from the legitimate site. Our system, on the other hand, had broad coverage and was able to identify phishing sites even when they did not have official logos.

\begin{figure}[!t]  
  \centering  
  \includegraphics[width=\linewidth]{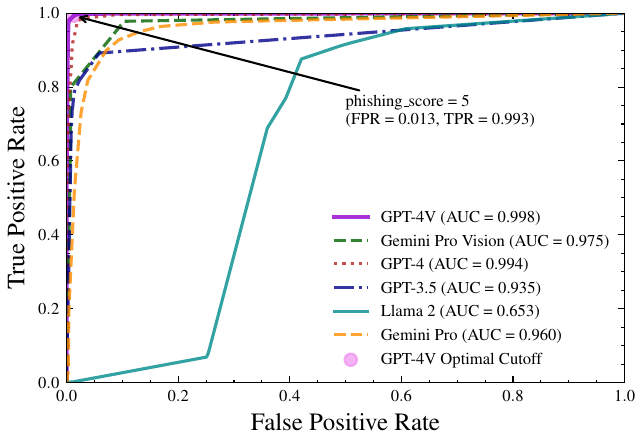}  
  \caption{ROC Curve for Phishing Classification Based on \textit{phishing\_score}.}  
  \label{fig:roc}  
\end{figure}

\noindent\textbf{Phishing Classification Using phishing\_score}
The above results were calculated based on the values of the \textit{phishing} and \textit{suspicious\_domain} keys, we can also classify the responses by selecting an appropriate threshold for the \textit{phishing\_score} value.
By incrementing the threshold by 1, we calculated the true positive rate and false positive rate and plotted the receiver operating characteristics (ROC) curve, as illustrated in Figure~\ref{fig:roc}. The optimal cutoff values for \textit{phishing\_score}, calculated using the Youden's J statistic ($ J = \frac{{TP}}{{TP + FN}} + \frac{{TN}}{{TN + FP}} - 1 $), were 5, 1, 3, 1, 6, and 4 for GPT-4V, Gemini Pro Vision, GPT-4, GPT-3.5, Llama2, and Gemini Pro.
The detection accuracy based on the threshold setting for \textit{phishing\_score} also demonstrated the superior performance of GPT-4V compared to other models.

\subsection{Cost Analysis and Performance Overhead}
In this section, we describe the cost and performance overhead of using ChatGPT's API for analysis with our system.

\noindent\textbf{Financial Cost}
In our evaluation experiment using the ChatGPT API for our dataset of 2,000 websites, the total costs for GPT-4V, GPT-4, and GPT-3.5 were \$358.65, \$304.26, and \$16.30, respectively.
As a result, the average cost per website was \$0.179 （GPT-4V), \$0.152 (GPT-4), and \$0.008 (GPT-3.5).

\noindent\textbf{Performance Overhead}
In terms of response time, the average duration for sending a request and receiving a response from ChatGPT was 25.13 seconds for GPT-4V, 18.19 seconds for GPT-4, and 14.42 seconds for GPT-3.5.

These results are significant in the current state of phishing countermeasures, which often involve extensive manual efforts in various domains, such as maintaining community-based platforms (e.g., PhishTank), creating threat intelligence by security vendors, and conducting law enforcement investigations. Our system not only ensures the identification of phishing sites with high accuracy but also delivers results rapidly and cost-effectively. Moreover, the system can generate reports, including detailed evidence of phishing sites written in different languages, in just a few tens of seconds, potentially surpassing the abilities of human experts. These aspects are very beneficial when our system is deployed in a real-world environment.

\subsection{Examples of LLM Responses}
\label{sec:example_response}

In this section, we describe specific examples of responses when LLMs correctly detect phishing sites.

\noindent\textbf{Phishing Example A}

\begin{figure}[!t]  
  \centering
    \shadowbox{\includegraphics[width=0.8\linewidth]{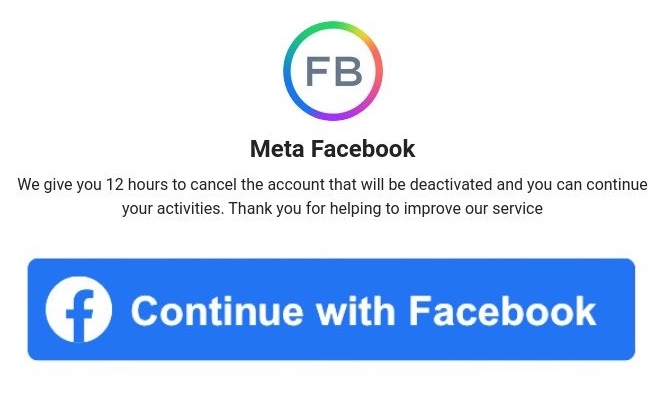}}
  \caption{Phishing Example A: Facebook Phishing Site.}
  \label{fig:example_phish_01}  
\end{figure}

\begin{tcolorbox}[title=Response 1 - GPT-4,breakable,fontupper=\small\ttfamily]
1. Suspicious elements:\\
- The URL contains a subdomain "securitypages", which is unrelated to the legitimate Facebook domain (\url{www.facebook[.]com}).\\
- The extracted text alerts the user about having 12 hours to cancel the account. This is a social engineering technique often used in phishing attacks.\\
- Furthermore, the link provided redirects to a suspicious, non-legitimate URL: "\url{http://centrenotificationvices.duckdns[.]org/cntre.html}".\\
2. The brand name identified is Meta Facebook.\\
3. The site appears to be a phishing site because of the suspicious URL and social engineering tactics in the text.\\
4. JSON-formatted output:
\begin{verbatim}
```
{
  "phishing_score": 9,
  "brands": "Meta Facebook",
  "phishing": true,
  "suspicious_domain": true
}
```
\end{verbatim}
\end{tcolorbox}

Figure~\ref{fig:example_phish_01} displays a screenshot image of a phishing site (\urlstyle{tt}{\url{securitypages.start[.]page}}), and Response 1 represents the response generated by GPT-4. We have defanged some of the URLs included in the response. This website impersonates Facebook and attempts to create a sense of urgency regarding account issues for the user. GPT-4 correctly identified this website as being associated with the ``Meta Facebook'' brand and determined that the domain name was not legitimate Facebook. Furthermore, GPT-4 recognized the presence of SE techniques and highlights the existence of suspicious links within the HTML. GPT-4 set the \textit{phishing\_score} to 9, marking both \textit{phishing} and \textit{suspicious\_domain} as true.

\noindent\textbf{Phishing Example B}

\begin{figure}[!t]  
  \centering  
  \shadowbox{\includegraphics[width=0.7\linewidth]{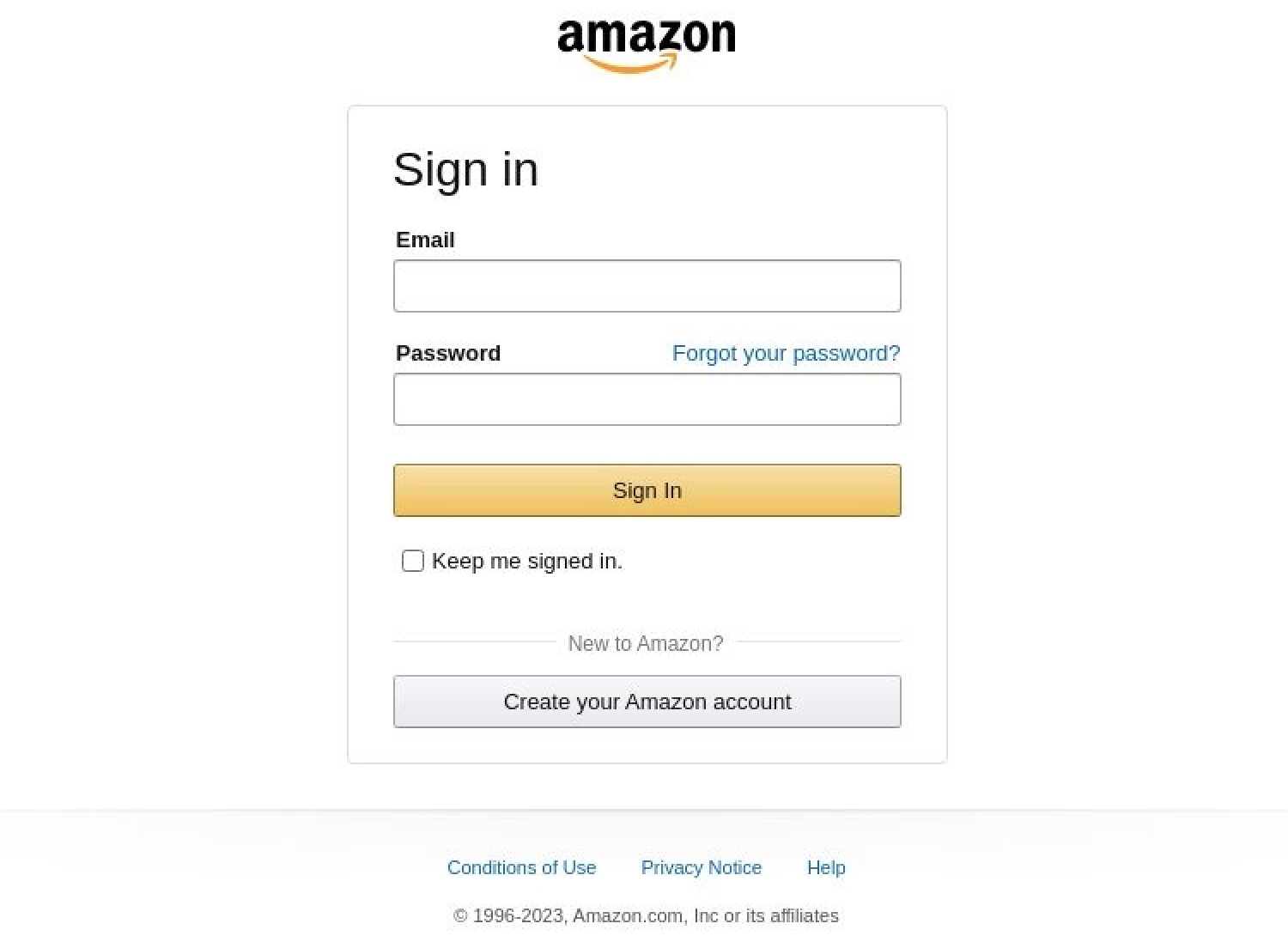}}
  \caption{Phishing Example B: Amazon Phishing Site.}  
  \label{fig:example_phish_02}  
\end{figure}  

\begin{tcolorbox}[title=Response 2 - GPT-4V,breakable,fontupper=\small\ttfamily]
1. Analyze the HTML, URL, and screenshot image:\\
- URL: The domain name in the URL is "jogeshchaudhuricollege[.]org" which does not match any known Amazon domain, a known brand, and does not seem relevant to an Amazon sign-in page. Also, the URL contains a suspiciously long and complex query string.\\
- HTML: The presented HTML attempts to replicate the Amazon sign-in page, including the Amazon logo and branding. However, the action attribute of the form element points to "check\_login.php", which is not typically how Amazon's legitimate login processes work.\\
- Image: The screenshot image resembles an Amazon sign-in page, which is a typical tactic used in phishing attempts to gain the trust of the user. The layout and design are very similar to the authentic Amazon site.\\
2. Identify the brand name:\\
- The HTML and image suggest that the site is imitating the Amazon sign-in page.\\
3. Conclusion:\\
- The site appears to be a phishing site. The URL does not correspond with the legitimate Amazon domain, and the HTML and image are mimicking an Amazon sign-in page to trick users into providing their credentials.\\
4. JSON-formatted output:
\begin{verbatim}
```json
{
  "phishing_score": 9,
  "brands": "Amazon",
  "phishing": true,
  "suspicious_domain": true
}
```
\end{verbatim}
\end{tcolorbox}

Figure~\ref{fig:example_phish_02} shows a phishing site impersonating the Amazon login page. In Response 2, which is the result generated by GPT-4V, the brand ``Amazon'' was identified in both the HTML and screenshot. GPT-4V found suspicious elements in the HTML related to the login process and recognized visual similarities with the legitimate Amazon website. GPT-4V successfully identified phishing sites based on several factors, including the suspiciousness of the URL query string and inconsistencies with the legitimate domain name.

\subsection{Evidence for Identifying Phishing Sites}
\label{sec:evidence_phishing}

In the analysis of phishing sites using LLMs, an important question arises: what evidence do LLMs rely on to accurately identify phishing sites? This section examines seven specific types of evidence that the GPT-4 and GPT-4V models, which produced the most accurate results, used to detect phishing sites.

\noindent\textbf{Legitimacy of Domain Names}
The most crucial evidence for identifying phishing sites is when the inspected website impersonates a legitimate brand but its domain name does not match the official domain name.
Our prompt instructs LLMs to extract the brand name from the HTML, screenshot image, or OCR-extracted text and match it with the legitimate domain name. 
This process involves more than simple named entity recognition; it relies on advanced analytics to determine and extract the most appropriate brand name from all the information on the website. In addition, in vision mode, LLMs can accurately extract brands based on known brand logos, text strings within images, and the overall appearance of the website.
Since LLMs have already been trained on various brands and their corresponding domain names, they can accurately detect inconsistencies between the extracted brand and its official domain name.
LLMs can also identify if the domain name included in the provided URL is a fake domain name (domain squatting~\cite{TianJ0Y018,KintisMLCGPNA17}) attempting to deceive users. For example, they correctly recognized that \urlstyle{tt}\url{disc0rd[.]pro}, which displayed a download page for Discord's installer, was not the legitimate domain name \urlstyle{tt}\url{discord[.]com}. 
Also, GPT-4V explained \url{www.na-amazon-creturns[.]com} as \textit{The URL contains ``amazon'' but with added dashes and extra characters that are not typical of legitimate Amazon URLs.}

\noindent\textbf{Fake Security Warning}
LLMs are also effective in detecting fake security alerts such as  malware infection, a common SE technique on the web. 
This extends beyond the simple detection of whether an infection warning is being displayed, as it enables the assessment of authenticity by examining suspicious grammar, unnatural language usage, and overstated warnings as indicators of potential deception.
For instance, the website \urlstyle{tt}\url{landiingpages[.]beauty}, which displayed a screen resembling Microsoft Windows Defender's virus scan, was analyzed to display fake security warnings such as \textit{このPCへのアクセスはセキュリティ上の理由でブロックされています (Access to this PC is blocked for security reasons)} and \textit{脅威を発見 (Threat found) - Trojan Spyware App: Ads.financetrack(1).exe}. 
Furthermore, the website \urlstyle{tt}\url{plfkuwyacu.duckdns[.]org}, which displayed the logo of mobile carrier SoftBank, was recognized for a series of social engineering tactics. 
GPT-4 accurately analyzed the psychological manipulation employed, which involves urging users to install \textit{SoftBankセキュリティ無料版アプリ (SoftBank Security App Free Edition)}, after they encounter a fake infection warning on this website that displays \textit{マルウェアが検出されました (Malware detected)}.

\noindent\textbf{Fake Account Issue}
Phishing sites often display warnings claiming that there are issues with the user's account, aiming to steal their login credentials. An example of this is the website \urlstyle{tt}\url{m4agence.web[.]app}, which was analyzed as a phishing site targeting the bank Société Générale. GPT-4 identified that it employed a SE technique as follows: \textit{The text extracted using OCR alerts the user to a problem with their account, which can be a typical technique of phishing attacks.}
Furthermore, it explained the presence of suspicious HTML elements on this website as follows: \textit{The HTML contains a link to a different domain than the one specified in the visible text: \url{"https://dev-sgwebnetauth.pantheonsite[.]io/wp-content/sysconnect"} instead of \url{"www.societegenerale[.]fr/synchronisation-agence."}}
LLMs have a high capability of HTML analysis, such as detecting inconsistencies between the text of the \texttt{a} element and the link specified in the \texttt{href}.

\noindent\textbf{Urgent Payment Request}
Phishing sites that impersonate courier services and generate alerts about package issues, while creating a sense of urgency for payment, are targeting brands worldwide. 
An example is \urlstyle{tt}\url{royalmail-online[.]com}, which disguised as Royal Mail and requested additional payment due to the overweight of the shipment. LLMs identified this as a phishing site based on the mismatch in the domain name and the presence of phrases such as \textit{Interrupted delivery} and \textit{Additional charges of 0.76£}, indicating the employment of SE techniques.
In the case of a phishing site (\urlstyle{tt}\url{www.caaarem[.]mx}) impersonating UPS, the LLMs detected the suspicious phrase \textit{790,45 HUF fizetése: Szállítási költség (Pay 790,45 HUF: Shipping cost)''}. Additionally, GPT-4 provided an explanation stating, \textit{The request for the user's credit card information, including card number, expiration date, and security code.} GPT-4 was able to analyze the HTML to gain a detailed understanding of the elements that encourage user input.
Moreover, \urlstyle{tt}\url{tlmcjohsvz.duckdns[.]org} was identified as a phishing site demanding payment for outstanding tax obligations. GPT-4 provided the following statement as evidence: 
\textit{The HTML contains alerting text that seems like a phishing attempt, such as ``差押最終通知'' (Final Attachment Notice), informing the user about an unpaid tax and urging them to pay using specific methods.}

\noindent\textbf{Fake Login Error}
As demonstrated in Section~\ref{sec:example_response}, some phishing sites display misleading error messages, such as indicating that the login credentials are wrong or alerting users that their credit card number is invalid, even when no input has been provided. These SE techniques aim to deceive users and extract sensitive information. For example, the website \urlstyle{tt}\url{www.interceptionbookingconfirmation[.]com}, pretending to be Booking.com, prompted users to input their credit card information, accompanied by an alert stating, \textit{Your credit card was marked as invalid.} GPT-4V explained as follows: \textit{The HTML contains a message designed to create a sense of urgency, informing the user that their credit card has been marked as invalid and requesting an update within 24 hours.}
Similarly, the website \urlstyle{tt}\url{mail.sikkimrajshree[.]in} was identified as impersonating the logistics company SF Express. Based on the displayed text on the website, GPT-4 categorized it as a phishing site, providing the following statement as evidence: \textit{The HTML content contains an error message '无效的用户名或密码' which translates to 'Invalid username or password', indicating that the user's account might have a problem, which can be a social engineering technique used in phishing attacks.}

\noindent\textbf{Fake Reward}
Phishing sites often employ deceptive tactics to capture users' interest. One common initial step is to offer fake rewards, such as monetary prizes, cryptocurrencies, gift cards, or smartphones.
For instance, the website \urlstyle{tt}\url{www.mobilegoodies4you[.]com} impersonates SFR, a telecommunication company, and displayed a fraudulent cash prize. GPT-4 identified the phrase \textit{Votre adresse IP a été tirée au sort et vous avez une chance de gagner un 460,00 € en espèces! (Your IP address has been entered into a draw for a chance to win €460.00 in cash!)} within the website as an example of an unexpected reward. 
Similarly, the phishing site \urlstyle{tt}\url{allesettlemenie[.]top}, which impersonates the e-commerce brand Allegro, was identified as using SE techniques to attract users' interest such as \textit{Sign up or log in today to shop and earn cash! Withdrawal to bank account (limited to one mobile number)}.
In the case of a phishing site \urlstyle{tt}\url{me7q1.vetedeb[.]xyz} targeting Sberbank, users are prompted to participate in a survey in exchange for cash. GPT-4 recognized the use of a cash reward to attract users' interest, as indicated by the statement: \textit{The text contains social engineering techniques such as enticing users with rewards ('ПОЛУЧИТЬ ВОЗНАГРАЖДЕНИЕ ДО 600 000 РУБ')}, which means ``get reward up to 600,000 RUB.''

\noindent\textbf{Authentication Request}
Phishing sites not only target users' login credentials but also attempt to obtain Two-Factor Authentication (2FA) information from them.
For instance, the website \urlstyle{tt}\url{demo.crustncakes[.]com} impersonates the payment service BenefitPay and deceived users into inputting an SMS code for the purpose of verifying a transaction. GPT-4V explained that this technique as follows: \textit{The HTML contains social engineering techniques such as alerting the user to a problem with an expired SMS code and prompting for immediate action, which is a tactic often used by phishing attacks to create a sense of urgency.}

\subsection{False Detection in GPT-4}
To assess the performance limitations of the proposed system and to gain insights for potential improvements, we analyzed the false positives and false negatives generated by GPT-4, which outperformed other models in normal mode. In the subsequent sections, we use GPT-4 as a benchmark for comparison against other models.

\subsubsection{False Positives (GPT-4)}
To explore the causes of false positives in LLMs, we analyzed 17 false positives for GPT-4 by classifying them into six factors based on the responses generated by the model.

\noindent\textbf{Misidentifying SE Techniques}
As discussed in Section~\ref{sec:evidence_phishing}, GPT-4 demonstrated precise identification of various SE techniques commonly used in phishing sites. However, it falsely flagged some non-phishing sites as phishing.
In our experiment, GPT-4 misidentified \urlstyle{tt}\url{cfspart.impots.gouv[.]fr} as containing SE techniques in an HTML element.
This element, initially hidden using the \url{display:none} property, triggers an alert message upon receiving incorrect input in the form.
GPT-4 correctly recognized \urlstyle{tt}\url{impots.gouv[.]fr} as a legitimate domain name. However, it misidentified this website as a distinct domain name due to the presence of the subdomain ``cfspart''. This suggests a potential misunderstanding of domain name structures by GPT-4.
Although there were other cases where GPT-4 reported that some legitimate sites contained elements using SE techniques, they were all classified as non-phishing because GPT-4 correctly verified that they matched legitimate domain names.

\noindent\textbf{Multiple Domain Names of Legitimate Brands}
GPT-4 sometimes misclassified legitimate brands that operate multiple domains as phishing sites. This occurred when GPT-4 could identify one or several of these domain names but did not have the knowledge of the specific domain name it was checking. Regarding this factor, GPT-4 produced four false positives for the following domain names: \urlstyle{tt}\url{m.botw[.]com} (another domain name for \urlstyle{tt}\url{bankofthewest[.]com}), \urlstyle{tt}\url{aplusfcu.balancepro[.]org} (\urlstyle{tt}\url{aplusfcu[.]org}), \urlstyle{tt}\url{hb.redlink.com[.]ar} (\urlstyle{tt}\url{bancodelapampa.com[.]ar}), and \urlstyle{tt}\url{cloudflare[.]net} (\urlstyle{tt}\url{cloudflare[.]com}). Additionally, it falsely flagged the law firm Pinsent Masons' website \urlstyle{tt}\url{www.aboutcookies[.]org} as phishing. This false positive occurred because the extracted brand name Pinsent Masons was not present in the domain name.

\noindent\textbf{Global Brands with Local Domain Names}
False positives can occur when global brands use local domain names specific to different countries. GPT-4 produced false positives for three domain names due to their mismatch with the most recognized legitimate domain names: \urlstyle{tt}\url{www.aliexpress[.]us} (the most recognized domain name by GPT-4: \urlstyle{tt}\url{www.aliexpress[.]com}), \urlstyle{tt}\url{sube.garantibbva.com[.]tr} (\urlstyle{tt}\url{garantibbva[.]com}), and \urlstyle{tt}\url{www.sicredi.com[.]br} (\urlstyle{tt}\url{www.sicredi[.]com}).

\noindent\textbf{Numerous Subdomains}
GPT-4 sometimes failed to identify a legitimate domain name if it had multiple subdomains, even if the domain name was under a legitimate domain name known to GPT-4.
The following domain names were false positives in this category: \urlstyle{tt}\url{cloud.walletconnect[.]com}, \urlstyle{tt}\url{entry11.bk.mufg[.]jp}, and \urlstyle{tt}\url{www.my.commbank.com[.]au}.

\noindent\textbf{Non-English Websites}
While GPT-4 demonstrated high accuracy in identifying SE techniques across a wide range of languages, false positives can occur due to insufficient knowledge of non-English websites. The following two domain names were falsely flagged as they were not well-known: \urlstyle{tt}\url{justhost[.]ru} and \urlstyle{tt}\url{www.jb51[.]net}.

\noindent\textbf{Long URL Query String}
GPT-4 correctly recognized the domain name \urlstyle{tt}\url{auth.talktalk.co[.]uk} as legitimate and did not detect any SE techniques within the web page content. However, it falsely flagged this site as phishing due to an extremely long URL query string (360 characters) in the URL path.

LLMs are known to improve their performance through Retrieval Augmented Generation (RAG), which involves referencing external data. In this study, we employed LLMs solely as classifiers for identifying phishing sites. To further reduce false positives, this could be achieved by creating URL allow lists and utilizing them as external data references.
Additionally, this has the potential to effectively identify phishing sites by detecting discrepancies between these sites and legitimate domain names, thereby also contributing to the reduction of false negatives.

\subsubsection{False Negatives (GPT-4)} 
GPT-4 demonstrated an improvement in reducing false negatives compared to GPT-3.5, achieving an 11.7\% reduction. However, there are cases where GPT-4 erroneously classifies certain phishing sites as non-phishing (16 cases), necessitating a thorough analysis of the underlying causes.
Out of the 16 FNs, except for 2 cases, GPT-4 correctly identified the brand names of the remaining websites. Among these FNs, 8 websites lacked any descriptive text other than JSON-formatted content in their responses, making it impossible to analyze the basis of their classification.
There were 4 FNs where GPT-3.5 correctly identified the phishing sites, but GPT-4 did not.

\noindent\textbf{Falsely Identified Domain Names as Legitimate}
Despite correctly identifying the targeted brand names for the following 5 phishing sites, GPT-4 erroneously classified the domain names as legitimate. 
For instance, GPT-4 misclassified the phishing site \urlstyle{tt}\url{www.phototan-push[.]de} (disguised as \urlstyle{tt}\url{commerzbank[.]com}) by stating \textit{there's no suspicion surrounding the domain name}, thus incorrectly labeling it as a non-phishing site. 
The phishing site \urlstyle{tt}\url{www.gmenduroracingteam[.]it} was identified as non-phishing by GPT-4, because the legitimate URL of the brand (\urlstyle{tt}\url{https://www.mooney[.]it/}) that this phishing site deceived was included as a link in the HTML, even though the actual domain name is different.
The following three sites were identified as false negatives by GPT-4, but correctly identified by GPT-3.5.
A website impersonating \urlstyle{tt}\url{www.gov[.]uk} (\urlstyle{tt}\url{claim.redundancy-payments.org[.]uk}) was classified as phishing by GPT-3.5 due to its SE technique of requesting the national insurance numbers and bank details. In contrast, GPT-4 accurately identified the brand as ``GOV.UK'' but indicated \textit{it is likely that the site is a legitimate one}. GPT-4 also suggested \textit{it is always better to verify the authenticity of the URL by visiting the official government website}.
\urlstyle{tt}\url{etmmetaverse[.]com}, a phishing site for Office 365, was correctly classified as phishing by GPT-3.5 based on the mismatch in the domain name and links to other pages. However, GPT-4 offered a neutral explanation, stating \textit{there is not enough evidence supporting that this is a phishing site}, despite assigning a \textit{phishing\_score} of 5 and labeling both \textit{phishing} and \textit{suspicious\_domain} as \textit{unknown}.
Furthermore, both GPT-3.5 and GPT-4 correctly identified the brand SI-DEP (France's national Covid-19 screening system) for the phishing site \urlstyle{tt}\url{ca8567c7fa0141658f08b0dabe13d5ee.v1.radwarecloud.net}. While GPT-3.5 correctly pointed out \textit{The domain name "radwarecloud.net" does not appear to be associated with the brand name "SI-DEP"}, GPT-4 failed to detect the phishing attempt.

\noindent\textbf{Failure to Identify Domain Squatting}
The website \urlstyle{tt}\url{www.americanexpressseguros[.]com} was a phishing site for American Express, offering insurance solicitations. Despite containing the phrase \textit{Recibe hasta \$2,000.00 M.N de bonificación al contratar (Receive up to \$2,000.00 M.N of bonus when contracting)}, GPT-4 failed to recognize the SE technique employed. Although it was a case of domain squatting, GPT-4 incorrectly identified it as a legitimate domain.

\noindent\textbf{Failure to Identify SE Techniques}
The phishing site \urlstyle{tt}\url{phpstack-197144-1061735.cloudwaysapps[.]com} presented an insurance survey, without specifically targeting any particular brand. Since it lacks prominent elements such as rewards or a sense of urgency, it was mistakenly classified as a legitimate site.
Similarly, the e-commerce site \urlstyle{tt}\url{lojanewgeneration[.]com} attempted to attract users with the phrase \textit{daily offers,} yet GPT-4 failed to identify the SE technique employed, leading to its classification as a legitimate site.

\subsection{Comparative Analysis: GPT-4 vs. GPT-4V}

This section provides an in-depth comparison of the performance differences between GPT-4 and GPT-4V, both of which have high phishing detection capabilities in normal and vision modes. 
GPT-4V effectively reduced both false positives (from 17 to 13) and false negatives (from 16 to 4) when compared to GPT-4. 
A notable strength of GPT-4V is its ability to identify various SE techniques used to deceive users.
This capability is based on visual information, including logos and page layouts in screenshot images, in addition to the advanced context interpretation abilities found in GPT-4. 
GPT-4V can accurately analyze the legitimacy of content by assessing factors such as brand recognition from logo images and the similarity of page appearances to legitimate sites.
In its brand identification, GPT-4V identified logo images in 242 phishing sites and used them as one of the classification clues.
As examples of cases where GPT-4V correctly identified phishing sites and GPT-4 failed, GPT-4V noted that for \url{www.phototan-push[.]de}, \textit{The design of the page closely mimics that of the legitimate Commerzbank login page}.
Also, GPT-4V identified a phishing site at \url{www.gmenduroracingteam[.]it} (FN of GPT-4) by recognizing the brand logo of Mooney (a payment service) and detecting inconsistencies in the domain name. 
Additionally, for \url{claim.redundancy-payments.org[.]uk}, GPT-4V analyzed the screenshot image and correctly identified it as phishing, stating that \textit{The GOV.UK logo and design are used.} In the following, we analyzed the false positives and false negatives of GPT-4V.

\subsubsection{False Positives (GPT-4V)}
GPT-4V had 13 false positives, with 3 cases remaining unexplained due to the lack of descriptions other than JSON, and 2 cases were incorrectly detected for the same reasons as GPT-4. Notably, GPT-4V correctly identified 15 out of 17 false positives from GPT-4 as non-phishing.
GPT-4V incorrectly identified \url{country.db[.]com} as a phishing site, primarily because of an unfamiliar subdomain. However, the website received a low \textit{phishing\_score} of 3, mainly due to insufficient clear evidence supporting the phishing classification.
The website \url{www.kinopoisk[.]ru} was also incorrectly detected due to the presence of a CAPTCHA in a screenshot. The website \url{ib.bri.co[.]id} was misclassified as phishing because GPT-4V interpreted a request to download a mobile app in the screenshot as a malware download attempt. The Trade Desk's ``Control Over Your Personal Information'' page (\url{adsrvr[.]org}), was incorrectly flagged as a phishing site due to its different domain name from the legitimate \url{thetradedesk[.]com}. The inclusion of subdomains led to the misclassification of \url{involta.service-now[.]com} and \url{kr.battle[.]net} as suspicious. Furthermore, \url{mega[.]io} was incorrectly identified as phishing. While GPT-4V recognized the cloud storage brand ``Mega'' from the website, it mistakenly considered the correct domain name to be \url{mega[.]nz}. Similarly, \url{cloudflare[.]net} was incorrectly detected due to a different TLD from the recognized legitimate domain \url{cloudflare[.]com}.

\subsubsection{False Negatives (GPT-4V)}
GPT-4V had the lowest number of false negatives of any model with only 4 cases, two of which were also false negatives on GPT-4 for the same reason (\url{lojanewgeneration[.]com} and \url{www.americanexpressseguros[.]com}). GPT-4V successfully identified 14 out of 16 false negatives from GPT-4 as non-phishing. 
One of the remaining two FNs is the Safra Bank phishing site (\url{secure.safrabankoffshore[.]com}).
GPT-4V explained that \textit{the domain 'safrabankoffshore.com' may raise suspicion as it includes 'offshore,' which could be a tactic to give the false impression of a special or exclusive service.}
Furthermore, GPT-4V misidentified the login page of this site as legitimate through image analysis, leading to the false negative.
GPT-4V recognized the GOV.UK brand from the phishing site \url{bankruptcy-form-fn-prod-ods.insolvency-development.co[.]uk}, however, GPT-4V was unable to determine that the domain name was not legitimate.

\subsection{Comparative Analysis: GPT-4 vs. GPT-3.5}
To analyze the impact of differences in model performance on phishing site detection, we conducted a comparative analysis between two models in normal mode, GPT-4 and GPT-3.5. Specifically, we thoroughly examined phishing sites that were successfully detected by GPT-4 but not by GPT-3.5. This allows us to uncover the reasons why LLMs are successful in detecting phishing sites.
Among the 133 FNs of GPT-3.5, a total of 121 phishing sites were correctly identified by GPT-4. Upon analyzing these 121 phishing sites, we discovered three abilities in which GPT-4 exhibited superior performance compared to GPT-3.5:%
\begin{itemize}
\item Determine the suspiciousness of domain names.
\item Identify SE techniques.
\item Detect phishing sites comprehensively by using multiple factors.
\end{itemize}

\noindent\textbf{Ability to Determine the Suspiciousness of Domain Names}
GPT-4 and GPT-3.5 differed in their ability to assess the legitimacy of domain names, especially to determine whether a domain name matches a legitimate domain name or exhibits suspicious characteristics.
For instance, when analyzing the OpenAI phishing site \urlstyle{tt}\url{openai-gpt-4[.]com}, GPT-3.5 classified it as non-phishing due to the presence of the term ``openai.'' In contrast, GPT-4 correctly identified it as phishing since it differed from the actual domain name, which is \urlstyle{tt}\url{openai[.]com}.
In the case of a phishing site (\urlstyle{tt}\url{krakken-logi.mystrikingly[.]com}) impersonating the cryptocurrency exchange Kraken, both GPT-4 and GPT-3.5 successfully recognized the brand name. While GPT-4 identified the misspelling of ``\url{krakken-logi}'' as suspicious, GPT-3.5 erroneously classified it as a legitimate domain name based on the presence of ``\url{krakken}''.
Furthermore, both GPT-4 and GPT-3.5 accurately identified the phishing site \urlstyle{tt}\url{firstcitizncb[.]com} as distinct from the legitimate domain name \urlstyle{tt}\url{firstcitizens[.]com}. However, GPT-3.5 considered it to be a variable of the legitimate domain name and not a phishing site.
Another example is the phishing site \urlstyle{tt}\url{correos-es.firebaseapp[.]com}, which impersonates Spain's national postal service, Correos. While GPT-3.5 classified it as a legitimate domain name, GPT-4 correctly recognized it as a suspicious domain due to its hosting on Google Firebase.

\noindent\textbf{Ability to Identify SE Techniques}
GPT-4 successfully identified SE techniques that were missed by GPT-3.5 in some phishing sites. For example, in the case of \urlstyle{tt}\url{verify.vodafone-uk[.]com}, which GPT-3.5 explained that \textit{The page warns users about the potential termination of their phone number}. However, GPT-4 recognized the presence of SE techniques by explaining that the websites warned users about the potential termination of their phone number. Additionally, GPT-4 noted that \textit{The domain name in the URL is "verify.vodafone-uk.com", which seems odd with the hyphen between "vodafone" and "uk".}
In the phishing site (\urlstyle{tt}\url{aktivere-bnkid.web[.]app}) impersonating the Swedish identification system BankID, GPT-4 identified the SE technique by highlighting that it requested personal information such as the user's personal number, phone number, and user-ID. In contrast, GPT-3.5 incorrectly states that the website \textit{does not contain any obvious social engineering techniques}.

\noindent\textbf{Ability to Detect Phishing Sites Comprehensively by Using Multiple Factors}
In some cases, LLMs may mistakenly identify suspicious elements on legitimate sites.
For example, they identified legitimate error messages (e.g., ``wrong password'') displayed after submitting a form as an indication of social engineering. 
LLMs also found the inclusion of the year 2023 in the copyright notice to be suspect because of the future year. This is due to a lack of knowledge regarding the current year. However, GPT-4 tends to make overall accurate determinations by prioritizing more reliable information, although it occasionally provides incorrect evidence. On the other hand, GPT-3.5 may classify a phishing site as benign even though it identifies the correct pieces of evidence.
The following are examples where GPT-3.5 incorrectly identified as non-phishing that GPT-4 correctly detected.
The website \urlstyle{tt}\url{b2meguy.com} displayed an attention-grabbing statement at the top of the website: \textit{Obtenga acceso instantáneo al software de Quantum Al, y gane de 3200€ al día!" (Get instant access to Quantum Al's software, and earn from 3200€ per day!)}. Both GPT-4 and GPT-3.5 identified this statement as suspicious, however, while GPT-3.5 did not classify it as a phishing site, GPT-4 correctly identified it as phishing based on the explanation that \textit{it presents unrealistic promises of gains.}
Another example is the phishing site \urlstyle{tt}\url{verification-appeal-code.firebaseapp[.]com}, which pretended to be Facebook and asked users to respond to claims of intellectual property infringement. Both GPT-4 and GPT-3.5 detected the request for sensitive information from visitors, however, only GPT-4 flagged it as a phishing site based on the suspiciousness of the domain name.

\section{Limitations}
LLMs provide flexible and varied responses because its output is determined probabilistically. 
However, this also means that detection results can change depending on the experiment. 
In this study we used default values for LLMs parameters such as top\_p and temperature. In order to obtain reliable results for classification, adjustments for specific purposes may be required.

Methods that use LLMs, including our system, may be susceptible to prompt injection attacks~\cite{perez2022ignore,greshake2023youve}, where the original prompt is overwritten and malicious content is inserted into the response. Our system employs simple countermeasures by simplifying the HTML and clarifying text sections through the placement of triple backticks.
However,  phishing sites may use prompt injections in the future to avoid analysis by LLMs. More advanced defenses against prompt injection attacks will be needed.

Since each LLM is trained on data up to a specific time period (e.g., the ChatGPT models used in this study was trained on data through September 2021), it may not be able to accurately classify phishing and non-phishing sites associated with services created after that date. The classification ability could potentially be improved by externally referencing a list of domain names corresponding to brands or by using a fine-tuned LLM.

\section{Related Work}

An effective method for detecting phishing sites is to compare the appearance of a website with that of a legitimate one. 
Previous studies such as Abdelnabi et al.~\cite{AbdelnabiKF20}, Lin et al.~\cite{LinLDNCLSZD21}, and Liu et al.~\cite{Liu0YNDD22} detect the abuse of logo images by identifying and comparing them with legitimate ones, or extract features from the overall appearance of web pages to analyze their similarity.
While these deep learning-based methods are effective in detecting phishing sites created by copying HTML or reusing logos, they cannot detect phishing sites that use their own logos or do not impersonate brands. 
Our system aims to identify phishing sites that use SE techniques by analyzing the context of the web page, allowing for the detection of different types of phishing sites, not limited to the misuse of branding.

Methods have been proposed to detect domain squatting, where attackers obtain domain names similar to legitimate service domain names. Nikiforakis et al.~\cite{NikiforakisAMDPJ13}, Quinkert et al.~\cite{QuinkertDBH20}, Kintis et al.~\cite{KintisMLCGPNA17}, Agten et al.~\cite{AgtenJPN15}, and dnstwist~\cite{dnstwist} have proposed techniques that use rule-based approaches or machine learning models to identify malicious domain names by considering various factors such as minor variations, the inclusion of brand names, and changes in top-level domains.
ChatGPT showed high capabilities against domain squatting, as demonstrated in our evaluation experiments, by detecting minor character differences and fake domain names containing brand names.
There are also methods that use TLS certificates to detect phishing sites. Kim et al.\cite{KimCKDSAD21}, Bijmans et al.\cite{BijmansBSNW21}, and Drichel et al.~\cite{DrichelDBM21} have proposed such approaches that use machine learning algorithms to analyze features derived from TLS certificates or certificate transparency logs to detect phishing sites.

There have been attempts to apply LLMs to cybersecurity. One such example is VirusTotal Code Insight~\cite{vtinsight}, which is based on Google's security-focused LLM, Sec-PaLM. This model can explain the purpose and functionality of malware code in natural language, allowing security experts to gain a general understanding of the code's intentions.
An article~\cite{securelist} analyzed phishing URLs using ChatGPT with a simple prompt. The experiment showed a detection rate of 87.2\% and a false positive rate of 23.2\%, indicating that while it can block phishing sites, it can also mistakenly block legitimate sites. In contrast, our system includes not only the URL, but also HTML, text extracted from screenshots using OCR, and carefully designed prompts to achieve superior detection accuracy.
Roy et al.~\cite{roy2023generating} explored the ability of ChatGPT to automatically generate phishing sites. Their findings suggest that these generated phishing sites can mimic popular brands and employ various evasion tactics to evade anti-phishing systems.
It is increasingly important to implement security measures using LLMs to mitigate automated phishing campaigns.

\section{Conclusion}

In this paper, we proposed a novel system called ~\textsc{ChatPhishDetector} for detecting phishing sites using LLMs. 
Our system leverages a combination of web crawling techniques and contextual understanding of LLMs to automatically classify websites as phishing or not.
Through an evaluation experiment, our system achieved remarkable performance, demonstrating the ability of LLMs to efficiently detect phishing sites by identifying brand impersonation and social engineering.
Comparative analysis with different modes, various models, and existing phishing detection systems revealed that our system, using GPT-4V, has superior detection ability.
This research suggested important insights for utilizing LLMs in automated cybersecurity, enhancing their ability to analyze malicious web content.
We provide the dataset and experimental results only upon request from researchers due to commercially licensed data inclusion and copyright issues.

\bibliographystyle{ACM-Reference-Format}
\bibliography{bib}

\newpage
\appendix

\section{Simple Mode Prompt}
\label{appendix_template}
Prompt Template 2 is the simple mode prompt, as described in Section~\ref{sec:experimental_setup}, which takes only the URL as its input.

\begin{figure}[h]
\begin{prompt}{}{}
You are a web programmer and security expert tasked with examining a URL to determine if it is a phishing site or a legitimate site.\\
Submit your findings as JSON-formatted output with the following keys:\\
- phishing\_score: int (indicates phishing risk on a scale of 0 to 10)\\
- brands: str (identified brand name or None if not applicable)\\
- phishing: boolean (whether the site is a phishing site or a legitimate site)\\
- suspicious\_domain: boolean (whether the domain name is suspected to be not legitimate)\\
\\
URL:\\
\{URL\}
\end{prompt}
\end{figure}

\section{Detection Results Using \textsc{ChatPhishDetector}}
\label{sec:appendix_detection_results}

Table~\ref{tab:chatphishdetector_result} shows the detailed detection results of \textsc{ChatPhishDetector} for each mode and model. 
In our dataset, GPT-4V failed to provide responses for five phishing sites and one non-phishing site. This occurred because our prompts triggered Azure OpenAI's content management policy.
Although we successfully disabled content filters in the API to allow GPT-4 and GPT-3.5 to process all of our prompts, it appears that the disabling setting did not effectively work for GPT-4V.

\begin{table}[!]
    \centering
    \caption{Detection Results by Mode and Model for \textsc{ChatPhishDetector}. TP, TN, FP, FN Represent True Positives, True Negatives, False Positives, and False Negatives, Respectively.}
    \label{tab:chatphishdetector_result}
    \begin{tabular}{llrrrr}
    \toprule
        \textbf{Mode} & \textbf{Model} & \textbf{TP} & \textbf{TN} & \textbf{FP} & \textbf{FN}  \\ 
        \midrule
        Vision & GPT-4V & 991 & 986 & 13 & 4  \\ 
         & Gemini Pro Vision & 789 & 993 & 7 & 211  \\ 
        Normal & GPT-4 & 984 & 983 & 17 & 16  \\ 
         & GPT-3.5 & 867 & 985 & 15 & 133  \\ 
         & Llama-2-70B & 664 & 817 & 183 & 336  \\ 
         & Gemini Pro & 905 & 958 & 42 & 95  \\ 
        Simple & GPT-4 & 755 & 988 & 12 & 245  \\ 
         & GPT-3.5 & 775 & 989 & 11 & 225 \\ 
         \bottomrule
    \end{tabular}
\end{table}

\section{Phishing Brands}
\label{appendix_brands}
Table~\ref{tab:legitimate_brands} is a list of 147 brands that were targeted by phishing sites in our dataset.

\newcolumntype{L}[1]{>{\raggedright\arraybackslash}p{#1}}

\onecolumn
\begin{table*}[h]
    \centering
    \caption{Brands Targeted by Phishing Sites.}
    \label{tab:legitimate_brands}
\begin{longtable}{L{0.3\linewidth}L{0.3\linewidth}L{0.3\linewidth}}
\toprule
A+ Federal Credit Union & AEON CARD & Alpha Web \\
Amazon & American Express & Apple \\
Ardoiz & BECU Online Banking & BNL \\
BRImo & Banca Sella & Banco Desio \\
Banco La Pampa & Bancoagrícola & Bancolombia \\
Bank of America & Barclays & Battle.net \\
Bendigo Bank & Bradesco & CAJA \\
CIMB Ni & Chunghwa Post Co. & Citizens Bank \\
CodeSquare & Coin Wallet & Colissimo \\
Commonwealth Bank & Correos & Credomatic \\
Crypto.com & DHL & DKB \\
Deutsche Bank & Deutsche Post DHL Group & Disney \\
Docomo & DocuSign & Dropbox \\
EATA & EClick PORTAL & ETC \\
EXPRESSPAY & Ebay & Emirates Post \\
Entrust & Facebook & Fifth Third Momentum Banking \\
First Citizens Bank & GLOBAL PASS & GO online \\
GOV.UK & Garanti BBVA & Garena \\
Gazprom & Gemini & GitLab \\
Google & Gruppo BNP Paribas & IBC \\
ING & ImToken & Impots.gouv.fr \\
Instagram & Involta & JUHACHI-SHINWA BANK \\
JUNO & Kakao & Kusainon \\
La Banque Postale & Livelo & METAMASK \\
MIR VISA & M\&T Bank & Ma Banque \\
Mashreq Online Banking & MasterCard & Mercari \\
Microsoft & Minnesota Unemployment Insurance Benefits System & Mitsubishi UFJ Bank \\
Mitsui Sumitomo Card & Mygrow & NAB Internet Banking \\
NAVY FEDERAL Credit Union & NETFLIX & NLB Banka \\
Nexi & Nordea & ORLEN \\
OakPay & OneDrive & Orange \\
OurTime & Outlook.com & PNC Online Banking \\
PRESTÍA & PancakeSwap & PayPal \\
PayPay & Post canada & QUOTY \\
Qatar Post & RENNER & REVOLUT INTERNATIONAL BANK \\
Ronin Wallet & SAISONCARD & SBJ Bank \\
SDCCU & SFR & SMBC \\
SMBC Trust Bank & SPANKKI & STEAM \\
Santander & Satang & Schwab Safe \\
Scotiabank & Security Bank & SharePoint \\
Sicredi & Siemens & Slovenská POŠTA \\
Spotify & Swiss & Swiss Post \\
Swisscom & T-Mobile & TD Ameritrade \\
TESCO Bank & THEWEST & TalkTalk \\
Theta & Transcash & Trust Wallet \\
USAA & USPS Tracking & Ubisoft \\
UniCredit & Uphold & UPS \\
VALVE & VISA & Verizon \\
Vietcombank & WELLS FARGO & WalletConnect \\
WeTransfer & Weebly & Yahoo! \\
\bottomrule
\end{longtable}
\end{table*}
\clearpage
\twocolumn

\end{document}